\documentclass{SCIS2018}
\begin{document}

%%%%%%%%%%%%%%%%%%%%%%%%%%%%%%%%%%%%%%%%%%%%%%%%%%%%%%%
%%% Authors do not modify the information below
%%% 作者不需要修改此处信息
\ArticleType{RESEARCH PAPER}
%\SpecialTopic{}
\Year{2018}
\Month{}
\Vol{61}
\No{}
\DOI{}
\ArtNo{}
\ReceiveDate{}
\ReviseDate{}
\AcceptDate{}
\OnlineDate{}
%%%%%%%%%%%%%%%%%%%%%%%%%%%%%%%%%%%%%%%%%%%%%%%%%%%%%%%

%%% title: 标题
%%%   \title{title}{title for citation}
\title{Optimal Power Allocation for Secure Directional Modulation Networks with a Full-duplex UAV User}

%%% Corresponding author: 通信作者
%%%   \author[number]{Full name}{{email@xxx.com}}
%%% General author: 一般作者
%%%   \author[number]{Full name}{}
\author[1]{Feng SHU}{shufeng0101@163.com}
\author[1]{Zaoyu LU}{}
\author[2]{Shuo ZHANG}{}
\author[1]{Jin WANG}{}
\author[1]{Xiaobo ZHOU}{}
\author[1]{Linlin SUN}{}
\author[1]{\\Jinhui LU}{}
\author[2]{Jinyong LIN}{}
\author[2]{Wenlong Cai}{}
%%% Author information for page head. 页眉中的作者信息
\AuthorMark{Shu F}

%%% Authors for citation. 首页引用中的作者信息
\AuthorCitation{Shu F, Lu Z Y, Wang J, et al}

%%% Authors' contribution. 同等贡献
%\contributions{Authors A and B have the same contribution to this work.}

%%% Address. 地址
%%%   \address[number]{Affiliation, City {\rm Postcode}, Country}
\address[1]{School of Electronic and Optical Engineering, Nanjing University of Science and Technology, Nanjing 210094, China.}
\address[2]{Beijing Aerospace Automatic Control Institute, Beijing 100854}

%%% Abstract. 摘要
\abstract{This paper make an investigation of a secure unmanned aerial vehicle (UAV)-aided communication network based on directional modulation(DM), in which one ground base station (Alice), one legitimate full-duplex (FD) user (Bob) and one illegal receiver (Eve) are involved. In this network, Alice acts as a control center to transmit confidential message and artificial noise (AN). The UAV user, moving along a linear flight trajectory, is intended to receive the useful information from Alice. At the same time, it also sends AN signals to further interference Eve's channel.  Aiming at maximizing secrecy rate during the UAV flight process, a joint optimization problem is formulated corresponding to power allocation (PA) factors, beamforming vector, AN projection matrices. For simplicity, maximum ratio transmission, null-space projection and the leakage-based method are applied to form the transmit beamforming vector, AN projection matrix at Alice, and AN projection vector at Bob, respectively. Following this, the optimization problem reduces into a bivariate optimization programme with two PA factors. We put forward an alternating iterative algorithm to optimize the two PA factors. Simulation results demonstrate that the proposed strategy for FD mode achieves a higher SR than the half-duplex (HD) mode, and outperforms the FD mode with fixed PA strategy.}

%%% Keywords. 关键词

\maketitle

%%%%%%%%%%%%%%%%%%%%%%%%%%%%%%%%%%%%%%%%%%%%%%%%%%%%%%%
%%% The main text. 正文部分
%%%%%%%%%%%%%%%%%%%%%%%%%%%%%%%%%%%%%%%%%%%%%%%%%%%%%%%
\section{Introduction}
With the highly diversified application scenarios, unmanned aerial vehicles (UAVs)-aided wireless communications have a promising prospect in the coming future.\cite{Yong2016Wireless,JointWu2018,Li2018Mobile,Securezhang2017}. UAV has a variety of advantages, such as high mobility, flexible deployment, controllable trajectory and low cost\cite{Chen2018A,Yang2017Energy,Xu2018A}. The problems surrounding the UAV-aided networks mainly focus on trajectory optimization, user scheduling, resource allocation, energy harvesting, etc\cite{Cheng2018UAV,Zhang2018Trajectory,Zhou2018UAV,Xu2018A}. Besides, due to the broadcast characteristic of wireless signals, UAV-aided wireless communications are vulnerable to be hostilely attacked. Consequently, secure wireless transmission in UAV wireless networks is a very challenging issue at present and in the further\cite{Zhong2018Secure,Zhou2019,Cai2018Dual,Liu2017Secure}. A cooperative jamming approach in\cite{Zhong2018Secure} is proposed to secure the UAV communication through utilizing the neighbor UAVs as jammers to defend against the eavesdropper. For purpose of maximizing the minimum average secrecy rate, the authors in \cite{Zhou2019} leverage the alternating iterative algorithm and successive convex approximation technique to jointly optimize the trajectories and transmit powers of UAV base station/jammers. The authors of \cite{Cai2018Dual} investigate a UAV-enabled secure communication system that the UAV trajectories and user scheduling are jointly adjusted to maximize the minimum worst-case secrecy rate among the users within each period. In \cite{Liu2017Secure}, the authors optimize the power allocation (PA) strategy by combining the transmission outage probability and secrecy outage probability as a main performance metric.

Directional modulation (DM), as a key physical layer security (PLS) technology in wireless communication, has attracted a lot of attention from both academia and industry\cite{Babakhani2008Transmitter,Chen2017A,Zou2015Relay}. To further protect the security of UAV communication network, DM technology has been increasingly applied to UAV network due to the line-of-sight (LoS) aerial-ground link\cite{Yong2016Wireless,Lee2018UAV,Khuwaja2018A}. Artificial noise (AN) is usually utilized to disturb the potential eavesdroppers to improve the PLS in DM networks\cite{Negi2005Secret,Wu2017Secure}. The authors in \cite{Ding2014} project the orthogonal AN on the null-space of the steering vector along the intended direction to enhance the PLS. For the multicast scenarios, both the useful precoding vector and the AN projection matrix can be designed according to the leakage-based criterion\cite{Shu2017Artificial}. In \cite{Wan2018Power}, the authors maximize the secrecy rate by the alternating iterative PA strategy, which achieves a significant SR gain in the case of small-scale antenna array. In DM systems, the direction of arrival (DOA) needs to be estimated in advance for the DM synthesis. Two phase alignment methods are proposed to estimate DOA based on the parametric method in\cite{Feng2017Low}. In \cite{Shu2018Robust}, the authors presented a robust beamforming scheme of combining main-lobe-integration and leakage to further improved the security in DM systems in the presence of direction angle measurement errors. A secure precise wireless transmission with low-complexity structure by random subcarrier selection (RSCS) is proposed in\cite{Feng2017Secure}.

To improve the spectral efficiency of next-generation wireless networks, full-duplex (FD) transmission has attracted fast-growing attentions\cite{Shafie2016Artificial,Lee2015Full,Chen2015Physical}. In \cite{Shafie2016Artificial}, a cooperative FD jammer is introduced to generate the precoded AN for guaranteeing the security of legitimate transmissions without knowing the eavesdropper's channel state information (CSI). The authors of \cite{Chen2015Physical} design a FD jamming relay network, in which the relay node transmits jamming signals to interfere the eavesdropper while receives the data from the source at the same time. FD transceivers are also adopted to enhance wireless PLS for multi-hop relaying systems in \cite{Lee2015Full}. When the relays receive information signals from the previous adjacent node, they simultaneously transmit AN signals to the eavesdropper.

Motivated by the above studies, in this paper, we investigate a PA strategy in UAV-aided communication networks where UAV user operates in FD mode. With the purpose of maximum secrecy rate (Max-SR), we formulate a joint optimization problem to design the beamforming vector, AN projection matrix, and PA factors. This problem is very hard to tackle due to the complicated objective function and the coupled variables. For simplicity, maximum ratio transmission (MRT) is applied to form the transmit beamforming vector, the AN projection matrix for Alice is constructed by the null-space projection (NSP) criterion, and the AN projection vector for Bob is designed by the leakage-based method. As such, the optimization problem reduces into a bivariate optimization programme with two PA factors. Using the given PA factor of the ground base station transmitter, we can design the optimal PA factor of legitimate UAV user. Correspondingly, with the given PA factor of UAV transmitter, we can derive the optimal PA factor of ground base station. Subsequently, an alternating iterative algorithm between two PA factors is proposed to further improve the SR performance. This algorithm is repeated until the terminal condition is satisfied. Simulation results also verify that the SR performance of the proposed strategy achieves a substantial gain over HD mode with fixed PA strategy.

The remainder of the paper is organized as follows. System model is described in Section II. Section III gives the beamforming vector and AN
projection vector, and an alternating iterative algorithm for PA strategy is proposed to further improve SR performance. The simulation and numerical results are shown in section IV. Finally, section V concludes this paper.

Notations: Throughout the paper, matrices, vectors, and scalars are denoted by letters of bold upper case, bold lower case, and lower case, respectively. Signs $(\cdot)^T$, $(\cdot)^H$, $\mid\cdot\mid$ and $\parallel\cdot\parallel$ represent transpose, conjugate transpose, modulus and norm, respectively. $\textbf{I}_N$ denotes the $N\times N$ identity matrix.

\section{SYSTEM MODEL}
\begin{figure}[!t]
  \centering
  \includegraphics[width=0.6\textwidth]{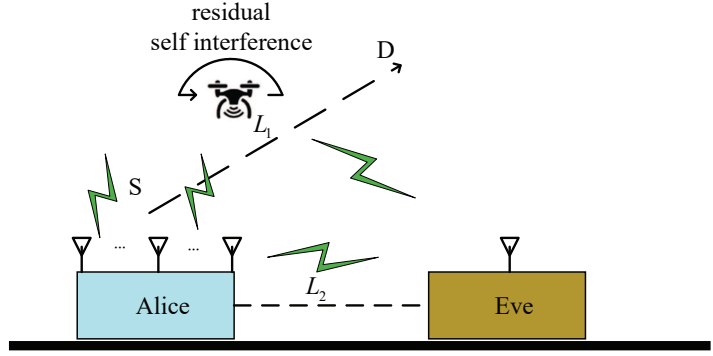}
  \caption{System model with FD receiver.}
  \label{Sys_Mod}
\end{figure}

Consider a UAV-enable wireless communications system as depicted in Fig.~\ref{Sys_Mod}. It consists of an $N$-antennas base station (Alice), a legitimate $N_b$-antennas UAV user (Bob) and a potentially illegal single-antenna receiver (Eve). Assuming that Bob flies along an $L_1$-meters-long direct path from S towards D. Bob works in FD mode, that is, it transmits signals with $N$ antennas while simultaneously receives messages with the remaining $N_b-N$ antennas. As mentioned in \cite{Lee2018UAV,Khuwaja2018A}, the link from ground station to the UAV user can be viewed as line-of-sight (LoS) channel, which means a single receive antenna is sufficient to provide available channel capacity. Hence, we set $N_b-N=1$.

Due to the broadcasting characteristic of wireless communications, the confidential messages conveyed from Alice to Bob are vulnerable to eavesdropping. Two measures are introduced to solve this problem. For the source node, AN signals are superimposed on the confidential messages to prevent the useful information being wiretapped. From the perspective of the destination node, Bob helps itself by operating in FD mode and transmitting AN signals to weaken the quality of Alice-to-Eve link. The transmitted baseband signal from Alice is expressed as

\begin{equation}\label{S}
\mathbf{s}=\sqrt{\beta_1P_a}\mathbf{v}_{b}x+\sqrt{(1-\beta_1)P_a}\mathbf{P}_{AN}\mathbf{z}
\end{equation}
where $P_a$ is the total transmit power at Alice, $\beta_1$ and $(1-\beta_1)$ stand for the PA factors for confidential message and AN, respectively. $x$ is the confidential message satisfying $\mathbb{E}\left\{x^Hx\right\}=1$, and $\mathbf{z}\in\mathbb{C}^{N\times1}$ means the AN vector obeying complex Gaussian distribution, i.e., $\mathbf{z}\sim\mathcal{C}\mathcal{N}(\mathbf{0},\mathbf{I}_{N})$. $\mathbf{v}_b\in\mathbb{C}^{N\times1}$ denotes the transmit beamforming vector for controlling the confidential messages to the desired direction, and $\mathbf{P}_{AN}\in\mathbb{C}^{N\times N} $ is the projection matrix leading AN to the undesired direction, $\mathbf{v}_b^H\mathbf{v}_b=1$ and $\mathrm{Tr}\{\mathbf{P}_{AN}\mathbf{P}_{AN}^H\}=1$.

The corresponding received signal at Bob can be written as
\begin{align}\label{Rx_signal yb}
y_{b}
=\sqrt{g_{ab}\beta_1P_a}\mathbf{h}^H_{sd}\mathbf{v}_{b}x
+\sqrt{g_{ab}(1-\beta_1)P_a}\mathbf{h}^H_{sd}\mathbf{P}_{AN}\mathbf{z}
+\sqrt{\beta_2P_b\rho}\mathbf{h}^H_{dd}\mathbf{q}_{AN}z+{n}_{b},
\end{align}
where $z\sim\mathcal{C}\mathcal{N}(0,1)$ denotes the scalar AN, and the complex additive white Gaussian noise (AWGN) at Bob is denoted by $n_{b}\sim\mathcal{C}\mathcal{N}(0,\sigma_b^2)$. $P_b$ represents the total transmit power at Bob. It is noted that if AN signal $z$ is sent at full power, it will generate a strong self-interference. Hence, $\beta_2$ is introduced to regulate the transmit power of AN signal. $g_{ab}=\frac{\alpha}{d_{ab}^c}$ represents the path loss from Alice to Bob. Here, $d_{ab}$ is the distance between Alice and Bob, $c$ denotes the path loss exponent and $\alpha$ means the path loss at reference distance $d_0$. $\mathbf{h}_{sd}\in\mathbb{C}^{N\times1}$ represents the steering vector from Alice to Bob, and $\mathbf{h}_{dd}\in\mathbb{C}^{N\times1}$ is the loop-channel between the receiving and the transmitting antennas of Bob. $\mathbf{q}_{AN}\in\mathbb{C}^{N\times1}$ stands for the projection matrix leading AN signal sent from Bob to the undesired direction with $\mathbf{q}_{AN}^H\mathbf{q}_{AN}=1$. $\rho \in[0,1]$ is introduced to represent the proportion of residual self-interference after antenna and radio-frequency cancellation.

In the same manner, the received signal at Eve is given by
\begin{align}\label{Rx_signal ye}
y_{e}
=\sqrt{g_{ae}\beta_1P_a}\mathbf{h}^H_{se}\mathbf{v}_{b}x
+\sqrt{g_{ae}(1-\beta_1)P_a}\mathbf{h}^H_{se}\mathbf{P}_{AN}\mathbf{z}
+\sqrt{g_{be}\beta_2P_b}\mathbf{h}^H_{de}\mathbf{q}_{AN}z+{n}_{e},
\end{align}
in which, $g_{ae}=\frac{\alpha}{d_{ae}^c}$ denotes the path loss from Alice to Eve, and $d_{ae}$ is the distance between Alice and Eve. $g_{be}=\frac{\alpha}{d_{be}^c}$ represents the path loss from Bob to Eve, and $d_{be}$ means the distance between Alice and Eve. $\mathbf{h}_{se}\in\mathbb{C}^{N\times1}$ denotes the steering vector from Alice to Eve, and $\mathbf{h}_{de}\in\mathbb{C}^{N\times1}$ denotes the steering vector from Bob to Eve. $n_{e}\sim\mathcal{C}\mathcal{N}(0,\sigma_e^2)$ is the complex AWGN at Eve.

For DM systems, the steering vectors in (\ref{Rx_signal yb}) and (\ref{Rx_signal ye}) have the following form
\begin{equation}\label{h_theta_b}
\mathbf{h}(\theta)=\frac{1}{\sqrt{N}}\left[e^{j2\pi\Psi_{\theta}(1)}, \cdots, e^{j2\pi\Psi_{\theta}(n)}, \cdots, e^{j2\pi\Psi_{\theta}(N)}\right]^T,
\end{equation}
and the phase function $\Psi_{\theta}(n)$ is defined by
\begin{equation}\label{var_phi}
\Psi_{\theta}(n)\triangleq-\frac{(n-(N+1)/2)d\cos\theta}{\lambda}, n=1,2,\cdots, N,
\end{equation}
where $n$ indexes the antenna, $d$ represents the antenna spacing, $\theta$ denotes the directional angle, and $\lambda$ means the carrier wavelength of transmitted signal.

As per (\ref{Rx_signal yb}) and (\ref{Rx_signal ye}), the achievable rates from Alice to Bob and Eve can be expressed as
\begin{align}\label{Rb}
R_{b}=\log_2
\left(1+\frac{g_{ab}P_a\beta_1\mathbf{h}^H_{sd}\tilde{\mathbf{V}}\mathbf{h}_{sd}}{g_{ab}(1-\beta_1)P_a \mathbf{h}^{H}_{sd}\tilde{\mathbf{P}}\mathbf{h}_{sd}+
\beta_2P_b\rho\mathbf{h}^{H}_{dd}\tilde{\mathbf{Q}}\mathbf{h}_{dd}+\sigma^2_b} \right),
\end{align}
and
\begin{align}\label{Re}
R_{e}=\log_2
\left(1+\frac{g_{ae}P_a\beta_1\mathbf{h}^H_{se}\tilde{\mathbf{V}}\mathbf{h}_{se}}{g_{ae}(1-\beta_1)P_a \mathbf{h}^{H}_{se}\tilde{\mathbf{P}}\mathbf{h}_{se}+
g_{be}\beta_2P_b\mathbf{h}^{H}_{de}\tilde{\mathbf{Q}}\mathbf{h}_{de}+\sigma^2_e} \right),
\end{align}

where
$\tilde{\mathbf{V}}=\mathbf{v}_{b}\mathbf{v}^H_{b}$, $\tilde{\mathbf{P}}=\mathbf{P}_{AN}\mathbf{P}^H_{AN}$ and $\tilde{\mathbf{Q}}=\mathbf{q}_{AN}\mathbf{q}^H_{AN}$.

As such, the achievable SR will become

\begin{align}\label{Rs}
R_s&=\max\left\{0,R_{b}-R_{e}\right\}
\end{align}

To maximize the above $R_s$, we need to solve the following optimization problem, that is
\begin{align}\label{P1}
\mathrm{(P1):}&\max_{\beta_1,\beta_2,\mathbf{v}_{b},\mathbf{P}_{AN},\mathbf{q}_{AN}}~~~~R_s\nonumber\\
&~~~~~~~~\text{s. t.}~~~~~~~0\leqslant\beta_1\leqslant1,\nonumber\\
&~~~~~~~~~~~~~~~~~~~~0\leqslant\beta_2\leqslant1,\nonumber\\
&~~~~~~~~~~~~~~~~~~~~\mathbf{v}^H_{b}\mathbf{v}_{b}=1,\nonumber\\
&~~~~~~~~~~~~~~~~~~~~\text{Tr}\{\mathbf{P}^H_{AN}\mathbf{P}_{AN}\}=1,\nonumber\\
&~~~~~~~~~~~~~~~~~~~~\mathbf{q}^H_{AN}\mathbf{q}_{AN}=1.
\end{align}

Obviously, this problem is difficult to tackle due to the complicated objective function. But, if the beamforming vector $\mathbf{v}_{b}$, $\mathbf{P}_{AN}$ and $\mathbf{q}_{AN}$ are designed or fixed in advance, the joint optimization will reduce into the following bivariate PA problem
\begin{align}\label{PA1}
\mathrm{(P1-1):}&\max_{\beta_1,\beta_2}~~~~R_s\nonumber\\
&~\text{s. t.}~~0\leqslant\beta_1\leqslant1,\nonumber\\
&~~~~~~~~0\leqslant\beta_2\leqslant1.
\end{align}
For simplicity, we rewrite $R_{b}-R_{e}$ as
\begin{align}\label{Rbe}
R_{b}-R_{e}=\log_2\left(1+\frac{A\beta_1}{(1-\beta_1)B+\beta_2C+\sigma^2_b}\right)
-\log_2\left(1+\frac{D\beta_1}{(1-\beta_1)E+\beta_2F+\sigma^2_e}\right),
\end{align}

where
\begin{align}
A&=g_{ab}P_a\mathbf{h}^H_{sd}\tilde{\mathbf{V}}\mathbf{h}_{sd},\\
B&=g_{ab}P_a\mathbf{h}^{H}_{sd}\tilde{\mathbf{P}}\mathbf{h}_{sd},\\
C&=P_b\rho\mathbf{h}^{H}_{dd}\tilde{\mathbf{Q}}\mathbf{h}_{dd},\\
D&=g_{ae}P_a\mathbf{h}^H_{se}\tilde{\mathbf{V}}\mathbf{h}_{se},\\
E&=g_{ae}P_a\mathbf{h}^{H}_{se}\tilde{\mathbf{P}}\mathbf{h}_{se},\\
F&=g_{be}P_b\mathbf{h}^{H}_{de}\tilde{\mathbf{Q}}\mathbf{h}_{de}.
\end{align}

Let $R_{b}-R_{e}=\emph{F}(\beta_1,\beta_2)$, optimization problem (\ref{PA1}) can be further reduced into
\begin{align}\label{PA2}
\mathrm{(P1-2):}&\max_{\beta_1,\beta_2}~~~~\max\{0,\emph{F}(\beta_1,\beta_2)\}\nonumber\\
&~~\text{s. t.}~~~~~0\leqslant\beta_1\leqslant1,\nonumber\\
&~~~~~~~~~~~~0\leqslant\beta_2\leqslant1.
\end{align}

For above objective function, the following equation holds
\begin{align}
\max_{\beta_1,\beta_2}~\max\{0,\emph{F}(\beta_1,\beta_2)\}=\max\{0,\max_{\beta_1,\beta_2}~\emph{F}(\beta_1,\beta_2)\}
\end{align}

Since $\emph{F}(\beta_1,\beta_2)=0$ is feasible at $\beta_1=0$, we have
\begin{align}
\max_{\beta_1,\beta_2} \emph{F}(\beta_1,\beta_2)\geq 0
\end{align}

Herein, the optimization problem (\ref{PA2}) turns to be
\begin{align}\label{PA3}
\mathrm{(P1-3):}&\max_{\beta_1,\beta_2}~~~~\emph{F}(\beta_1,\beta_2)\nonumber\\
&~~\text{s. t.}~~~~~0\leqslant\beta_1\leqslant1,\nonumber\\
&~~~~~~~~~~~~0\leqslant\beta_2\leqslant1.
\end{align}

In other words, the optimization problem (\ref{PA1}) is equivalent to
\begin{align}\label{P2}
&\mathrm{(PA1):}\max_{\beta_1,\beta_2}~~~\emph{F}(\beta_1,\beta_2)\nonumber\\
&=\log_2\frac{-A_2\beta_1^2+B_1\beta_2^2+C_2\beta_1\beta_2+D_2\beta_1+E_1\beta_2+F_1}
{-A_1\beta_1^2+B_1\beta_2^2+C_1\beta_1\beta_2+D_1\beta_1+E_1\beta_2+F_1}\nonumber\\
&~~~~~~~~~\text{s. t.}~~~~~0\leqslant\beta_1\leqslant1,\nonumber\\
&~~~~~~~~~~~~~~~~~~~0\leqslant\beta_2\leqslant1.
\end{align}
where
\begin{align}
A_1&=B(D-E),\\
B_1&=CF,\\
C_1&=C(D-E)-BF,\\
D_1&=(D-E)(B+\sigma^2_b)-B(E+\sigma^2_b),\\
E_1&=C(E+\sigma^2_e)+F(B+\sigma^2_b),\\
F_1&=(B+\sigma^2_b)(E+\sigma^2_e),\\
A_2&=E(A-B),\\
C_2&=F(A-B)-CE,\\
D_2&=(A-B)(E+\sigma^2_e)-E(B+\sigma^2_b),
\end{align}

Observing (\ref{P2}), the expression of function $\emph{F}(\beta_1,\beta_2)$ indicates that $R_s$ is continuous and differentiable with respect to $\beta_1$, $\beta_2$ in closed interval $[0,1]$. Hence, the promising optimal $\beta_1$ and $\beta_2$ must be either endpoints or stationary points. As such, we need to find out all the stationary points by vanishing the first-order derivative of $\emph{F}(\beta_1,\beta_2)$ with respect to $\beta_1$ and $\beta_2$. On this basis, the optimal $\beta_1$, $\beta_2$ will be selected by comparing the values of $\emph{F}(\beta_1,\beta_2)$ among the set of all candidate points to the critical numbers.

\section{PROPOSED PA STRATEGY OF MAX-SR}
In this section, we employ NSP scheme to guarantee that the legitimate UAV receiver will not be affected by the AN signal, while at the same time the potentially illegal eavesdropper will be seriously distorted. Besides, from the aspect of Bob, we expect that all the AN signal transmitted by Bob will interfere the quality of the Alice-to-Eve link and minimize AN signal leakage to itself. Then the PA factors are iteratively solved.

\subsection{Design $\mathbf{v}_{b}$, $\mathbf{P}_{AN}$ and $\mathbf{q}_{AN}$}
MRT is applied to form the transmit beamforming vector at Alice
\begin{equation}
\mathbf{v}_{b}=\mathbf{h}_{sd}.
\end{equation}

In order to make AN signal emitted by Alice free of any interference to Bob, $\mathbf{P}_{AN}$ is designed to project $\mathbf{z}$ into the null space of $\mathbf{h}^{H}_{sd}$. Therefore, the projection matrix $\mathbf{P}_{AN}$ is given by
\begin{equation}
\mathbf{P}_{AN}=\mathbf{I}_{N}-\mathbf{h}_{sd}\mathbf{h}_{sd}^{H}.
\end{equation}

To further interfere with Eve, Bob will also transmit AN signal, which generates a strong self-interference. Motivated by this, we design beamforming vector $\mathbf{q}_{AN}$ to minimize the AN signal leakage to Bob, called maximizing AN-to-leakage-and-noise ratio (ANLNR),
\begin{align}\label{P}
\mathrm{(P2):}&\max_{\mathbf{q}_{AN}}~~~~\mathrm{ANLNR}(\mathbf{q}_{AN})\nonumber\\
&~~\text{s. t.}~~~~\mathbf{q}^H_{AN}\mathbf{q}_{AN}=1,
\end{align}
where
\begin{align}
\mathrm{ANLNR}(\mathbf{q}_{AN})=\frac{\beta_2 P_b\mathbf{q}^H_{AN}\mathbf{h}_{de}\mathbf{h}^{H}_{de}\mathbf{q}_{AN}}
{\mathbf{q}^H_{AN}(\beta_2 P_b\mathbf{h}_{dd}\mathbf{h}^{H}_{dd}+\sigma_e^2\mathbf{I}_{N})\mathbf{q}_{AN}}.
\end{align}

Using the generalized Rayleigh-Ritz ratio theorem in \cite{Feng2011Ane}, the optimal $\mathbf{q}_{AN}$ can be obtained from the eigenvector corresponding to the largest eigen-value of the matrix
\begin{align}
[\beta_2 P_b\mathbf{h}_{dd}\mathbf{h}^{H}_{dd}+\sigma_e^2\mathbf{I}_{N}]^{-1}\mathbf{h}_{de}\mathbf{h}^{H}_{de}.
\end{align}

Notice that the above matrix is rank-one, we can obtain the closed-form solution to (\ref{P}) as
\begin{align}\label{slo_van}
\mathbf{q}_{AN}=\frac{[\beta_2\mathbf{h}_{dd}\mathbf{h}^H_{dd}P_b+\sigma_e^2 \mathbf{I}_{N}]^{-1}\mathbf{h}_{de}}{\|[\beta_2\mathbf{h}_{dd}\mathbf{h}^H_{dd}P_b+\sigma_e^2 \mathbf{I}_{N}]^{-1}\mathbf{h}_{de}\|_2}.
\end{align}

On this basis, we will detail the PA scheme to solve $\beta_1$ for fixed $\beta_2$ in what follows.

\subsection{Optimize $\beta_1$ for fixed $\beta_2$}
For fixed $\beta_2$, the stationary points with respect to $\beta_1$ should satisfy the following equation
\begin{align}\label{fbeta_1}
\frac{\partial \emph{F}(\beta_1, \text{fixed}~\beta_2)}{\partial\beta_1}&=
\frac{(A_1B_3-A_2D_3)\beta^2_1+2(A_1C_3-A_2C_3)\beta_1+B_3C_3-C_3D_3}
{(-A_1\beta_1^2+D_3\beta_1+C_3)^2}\nonumber\\
&=0,
\end{align}
where
\begin{align}
B_3&=D_2+C_2\beta_2,\\
C_3&=(E_1+B_1\beta_2)\beta_2+F_1,\\
D_3&=D_1+C_1\beta_2,
\end{align}
which yields
\begin{align}\label{beta_11}
\beta_{1,1}=\frac{-(A_1C_3-A_2C_3)+\sqrt{\Delta_1}}{A_1B_3-A_2D_3},
\end{align}
\begin{align}\label{beta_12}
\beta_{1,2}=\frac{-(A_1C_3-A_2C_3)-\sqrt{\Delta_1}}{A_1B_3-A_2D_3},
\end{align}
in which
\begin{equation}
\Delta_1=(A_1C_3-A_2C_3)^2-(A_1B_3-A_2D_3)(B_3C_3-C_3D_3).
\end{equation}

Considering $\beta_1\in[0,1]$, we obtain the candidate set for the critical numbers of function $\emph{F}(\beta_1, \text{fixed}~\beta_2)$ as
\begin{equation}
S_1(\text{fixed}~\beta_2)=\{0,\beta_{1,1},\beta_{1,2},1\}.
\end{equation}

In the following, we need to decide which $\beta_1$ is the optimal solution to maximize the function $\emph{F}(\beta_1, \text{fixed}~\beta_2)$ in the four candidate points. Obviously, $\beta_1=0$ means that there is no confidential messages transmitted to Bob. Thus, $\beta_1=0$ should be removed from the above candidate set. Now we only need to discuss the remaining three candidate points $\emph{F}(\beta_{1,1}, \text{fixed}~\beta_2)$, $\emph{F}(\beta_{1,2}, \text{fixed}~\beta_2)$ and $\emph{F}(1, \text{fixed}~\beta_2)$.

Below, let us discuss this issue in different cases of $\Delta_1$. First case is $\Delta_1<0$, then the two real roots $\beta_{1,1}$ and $\beta_{1,2}$ will not exist. On this basis, we need to check  the following three cases.\\
\textbf{Case 1.} $A_1B_3-A_2D_3>0$, $\emph{F}(\beta_1, \text{fixed}~\beta_2)$ is a monotonously increasing function and will achieve its maximum value at $\beta_1=1$. \\
\textbf{Case 2.} $A_1B_3-A_2D_3=0$, the stationary point is $\beta_{1,3}=\frac{C_3D_3-B_3C_3}{2(A_1C_3-A_2C_3)}$. We obtain the PA factor $\beta_1$ by comparing the value of $\emph{F}(\beta_{1,3}, \text{fixed}~\beta_2)$ and $\emph{F}(1, \text{fixed}~\beta_2)$.\\
\textbf{Case 3.} $A_1B_3-A_2D_3<0$, $\emph{F}(\beta_1, \text{fixed}~\beta_2)$ is a monotonously decreasing function. Therefore, the PA factor is $\beta_1=0$. This result leads to a contradict with what was discussed above.

The second case is $\Delta_1\geqslant0$, we need to judge whether $\beta_{1,1}$ and $\beta_{1,2}$ meet the condition that the PA factor $\beta_1$ lies in the interval of (0, 1). Then, compare the values of $\emph{F}(\beta_1, \text{fixed}~\beta_2)$ at the endpoints and corresponding stationary points to get the $\beta_1$. There are four different cases.\\
\textbf{Case 1.} If $\beta_{1,1}\in(0,1)$, $\beta_{1,2}\in(0,1)$, then compare the values of $\emph{F}(\beta_{1,1}, \text{fixed}~\beta_2)$, $\emph{F}(\beta_{1,2}, \text{fixed}~\beta_2)$ and $\emph{F}(1, \text{fixed}~\beta_2)$.\\
\textbf{Case 2.} If $\beta_{1,1}\in(0,1)$, $\beta_{1,2}\notin(0,1)$, then compare the values of $\emph{F}(\beta_{1,1}, \text{fixed}~\beta_2)$ and $\emph{F}(1, \text{fixed}~\beta_2)$.\\
\textbf{Case 3.} If $\beta_{1,1}\notin(0,1)$, $\beta_{1,2}\in(0,1)$, then compare the values of $\emph{F}(\beta_{1,2}, \text{fixed}~\beta_2)$ and $\emph{F}(1, \text{fixed}~\beta_2)$.\\
\textbf{Case 4.} If $\beta_{1,1}\notin(0,1)$, $\beta_{1,2}\notin(0,1)$, then the value of $\beta=1$ will be the solution.

After the above discussion, we can get the optimal transmit PA factor $\beta_1^*$ for Alice.

\subsection{Optimize $\beta_2$ for fixed $\beta_1$}
Similarly, the stationary points with respect to $\beta_2$ for fixed $\beta_1$ can be found out by
\begin{align}\label{fbeta_2}
\frac{\partial \emph{F}(\text{fixed}~\beta_1,\beta_2)}{\partial\beta_2}&=
\frac{(B_1D_4-B_1B_4)\beta^2_2+2(B_1E_4-B_1C_4)\beta_2+B_4E_4-C_4D_4}
{(B_1\beta_2^2+D_4\beta_2+E_4)^2}\nonumber\\
&=0,
\end{align}
where
\begin{align}
B_4&=E_1+C_2\beta_1,\\
C_4&=(D_2-A_2\beta_1)\beta_1+F_1,\\
D_4&=E_1+C_1\beta_1,\\
E_4&=(D_1-A_1\beta_1)\beta_1+F_1,
\end{align}
which yields
\begin{align}\label{beta_21}
\beta_{2,1}=\frac{-(B_1E_4-B_1C_4)+\sqrt{\Delta_2}}{B_1D_4-B_1B_4},
\end{align}
\begin{align}\label{beta_22}
\beta_{2,2}=\frac{-(B_1E_4-B_1C_4)-\sqrt{\Delta_2}}{B_1D_4-B_1B_4},
\end{align}
where
\begin{equation}
\Delta_2=(B_1E_4-B_1C_4)^2-(B_1D_4-B_1B_4)(B_4E_4-C_4D_4).
\end{equation}

Now, we obtain the candidate set of $\beta_2$ for the critical numbers of function $\emph{F}(\text{fixed}~\beta_1,\beta_2)$ as
\begin{align}
S_2(\text{fixed}~\beta_1)=\left\{0,\beta_{2,1}, \beta_{2,2}, 1\right\}.
\end{align}

From the aspect of legitimate UAV receiver, $\beta_2=0$ means that there is no AN sent from Bob. This case happens when the potentially illegal receiver Eve is so weak that Bob has no need to send AN signal to interfere with the eavesdropper. On the contrary, when the AN signal transmitted from Alice is fairly weak, $\beta_2=1$ may be suitable for Bob to mostly disturb Eve. Overall, there are four candidate points $\emph{F}(\text{fixed}~\beta_1, 0)$, $\emph{F}(\text{fixed}~\beta_1, \beta_{2,1})$, $\emph{F}(\text{fixed}~\beta_1, \beta_{2,2})$ and $\emph{F}(\text{fixed}~\beta_1, 1)$ that we need to judge.

In the same way with $\beta_1$, we discuss $\beta_{2,1}$, and $\beta_{2,2}$ in different cases of $\Delta_2$, the first case is $\Delta_2<0$.\\
\textbf{Case 1.} $B_1D_4-B_1B_4>0$, $\emph{F}(\text{fixed}~\beta_1,\beta_2)$ is a monotonously increasing function. It will achieve the maximum value at $\beta_2=1$. \\
\textbf{Case 2.} $B_1D_4-B_1B_4=0$, the stationary point is $\beta_{2,3}=\frac{C_4D_4-B_4E_4}{2(B_1E_4-B_1C_4)}$. We obtain the PA factor $\beta_2$ by comparing the value of $\emph{F}(\text{fixed}~\beta_1, 0)$, $\emph{F}(\text{fixed}~\beta_1, \beta_{2,3})$ and $\emph{F}(\text{fixed}~\beta_1, 1)$.\\
\textbf{Case 3.} $B_1D_4-B_1B_4<0$, $\emph{F}(\text{fixed}~\beta_1,\beta_2)$ is a monotonously decreasing function. Therefore, the PA factor is set as $\beta_2=0$.

For the second case  $\Delta_2\geqslant0$, we should judge whether the $\beta_{2,1}$ and $\beta_{2,2}$ meets the condition that the PA factor $\beta_2$ lies in the interval of (0, 1). Then, compare the values of $\emph{F}(\text{fixed}~\beta_1,\beta_2)$ at the endpoints and corresponding stationary points to get the $\beta_2$. There are also four different cases.\\
\textbf{Case 1.} If $\beta_{2,1}\in(0,1)$, $\beta_{2,2}\in(0,1)$, then compare the values of $\emph{F}(\text{fixed}~\beta_1, 0)$, $\emph{F}(\text{fixed}~\beta_1, \beta_{2,1})$, $\emph{F}(\text{fixed}~\beta_1, \beta_{2,2})$ and $\emph{F}(\text{fixed}~\beta_1, 1)$.\\
\textbf{Case 2.} If $\beta_{2,1}\in(0,1)$, $\beta_{2,2}\notin(0,1)$, then compare the values of $\emph{F}(\text{fixed}~\beta_1, 0)$, $\emph{F}(\text{fixed}~\beta_1, \beta_{2,1})$ and $\emph{F}(\text{fixed}~\beta_1, 1)$.\\
\textbf{Case 3.} If $\beta_{2,1}\notin(0,1)$, $\beta_{2,2}\in(0,1)$, then compare the values of $\emph{F}(\text{fixed}~\beta_1, 0)$, $\emph{F}(\text{fixed}~\beta_1, \beta_{2,2})$ and $\emph{F}(\text{fixed}~\beta_1, 1)$.\\
\textbf{Case 4.} If $\beta_{2,1}\notin(0,1)$, $\beta_{2,2}\notin(0,1)$, then compare the value of $\emph{F}(\text{fixed}~\beta_1, 0)$ and $\emph{F}(\text{fixed}~\beta_1, 1)$.

To this end, we can obtain the optimal transmit PA factor $\beta_2^*$ for the UAV user.

\subsection{Proposed alternating iterative algorithm}
After the above discussion, we can get the optimal PA factor $\beta_1$($\beta_2$) when $\beta_2$($\beta_1$) is given. Next, we propose an alternating iterative algorithm from the aspect of further improving SR performance. This alternating iterative algorithm is established between $\beta^i_1$ and $\beta^i_2$ with an initial value of $\beta^{0}_2$, where superscript $i$ indicates the $i$th iteration. Then, the feasible PA factor $\beta^0_1$ can be obtained from (\ref{fbeta_1}) and finally selected from three candidates via aforementioned two scenarios. Subsequently, with the value of $\beta^0_1$, we compute the values of $\beta^1_2$ based on (\ref{fbeta_2}) and select the optimal PA factor from four candidates as aforementioned. This process will be repeated until $|\emph{F}(\beta^i_1,\beta^{i}_2)-\emph{F}(\beta^{i-1}_1,\beta^{i-1}_2)|$ is smaller than a preset small value. To make clear, the iterative algorithm is summarized as Algorithm~\ref{algorithm 1}.

\begin{algorithm}
Initialization: $i=0, \beta^{i}_2=0.1, R_s^{i}=0$.
\begin{enumerate}
  \item For given $\beta^{i}_2$, solve (\ref{fbeta_1}) to obtain the two stationary points (\ref{beta_11}) and (\ref{beta_12}),
  \item Discuss aforementioned two scenarios to obtain optimal $\beta^{i}_1$,
  \item Solve (\ref{fbeta_2}) with the fixed $\beta^i_1$, and obtain the two stationary points (\ref{beta_21}) and (\ref{beta_22}),
  \item Discuss aforementioned two scenarios to obtain $\beta^{i+1}_2$,
  \item Update $\beta^{i}_2=\beta^{i+1}_2, i=i+1$,
  \item Compute $R_s^{i}$,
  \item Until $|\emph{F}(\beta^i_1,\beta^{i}_2)-\emph{F}(\beta^{i-1}_1,\beta^{i-1}_2)|\leq\epsilon$.
\end{enumerate}
\caption{Proposed alternating iterative algorithm}\label{algorithm 1}
\end{algorithm}

\section{Simulation and Discussion}
To evaluate the SR performance of the proposed strategy, simulation results are presented in this section. The system parameters are set as: the antenna spacing is $d=\lambda/2$, the path loss exponent $c=2$, the desired direction is $22^{\circ}$, undesired direction is $30^{\circ}$, the distance between $S$ and $D$ is $L_1$=400m, the distance between Alice and Eve is $L_2$=200m, the UAV speed is $v$=10m/s. To ensure fairness, the transmit power used in HD mode is the summation of the Alice itself and the actual transmit power of Bob.

\begin{figure}[!t]
\centering
\includegraphics[width=0.6\textwidth]{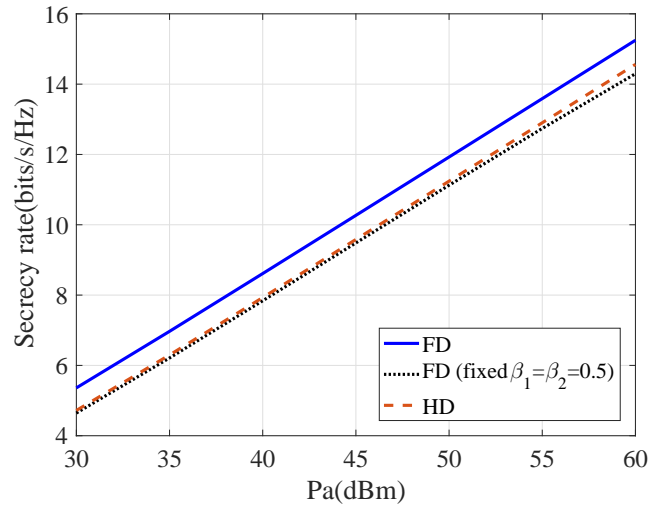}
\caption{SR versus $P_a$ for three methods.}
\label{SR_P}
\end{figure}

\begin{figure}[!t]
\centering
\includegraphics[width=0.6\textwidth]{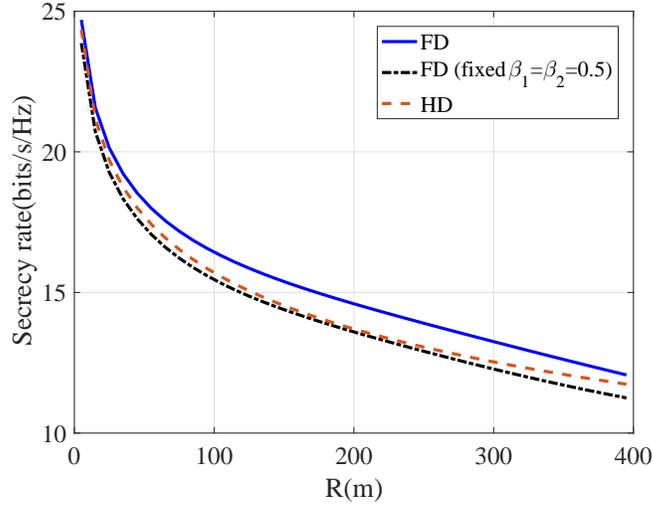}
\caption{SR versus $R$ for three methods.}
\label{SR_x}
\end{figure}

Fig.~\ref{SR_P} plots the curves of SR versus $P_a$ for the proposed PA strategy when the UAV user operates in FD mode and HD mode. For comparison, FD mode with fixed PA factor is also introduced. From Fig.~\ref{SR_P}, we can clearly see that the SR increases gradually with the transmit power $P_a$ no matter Bob operates in FD or HD mode. For any fixed transmit power $P_a$, it is manifest that the FD mode  achieves a substantial SR performance gain over the HD mode and shows a great improvements over the case of fixed PA factor. However, the HD mode achieves a little bit of SR performance gains over the FD mode with fixed PA factor $\beta_1=\beta_2=0.5$. It reveals that FD mode is not always more reliable than HD mode because of the existence of inevitable self-interference. This also verifies that, for the FD UAV user, our proposed PA strategy is valid to achieve a power balance between the efficient AN signal and redundant self-interference.

Fig.~\ref{SR_x} shows the curves of SR versus $R$ (the flight distance between Bob and Alice) for the proposed strategy when the UAV user operates in FD mode and HD mode. Also, FD mode with fixed PA factors is taken into consideration as $\beta_1=\beta_2=0.5$. From this figure, we can observe that the proposed strategy in FD mode shows a remarkable improvement over the case of HD mode. And PA factors generated by the proposed iterative algorithm is superior to the fixed case from the aspect of SR performance. Specifically, with Bob flying far away form Alice, the SR performance gain becomes more significant.

\begin{figure}[!t]
\centering
\includegraphics[width=0.6\textwidth]{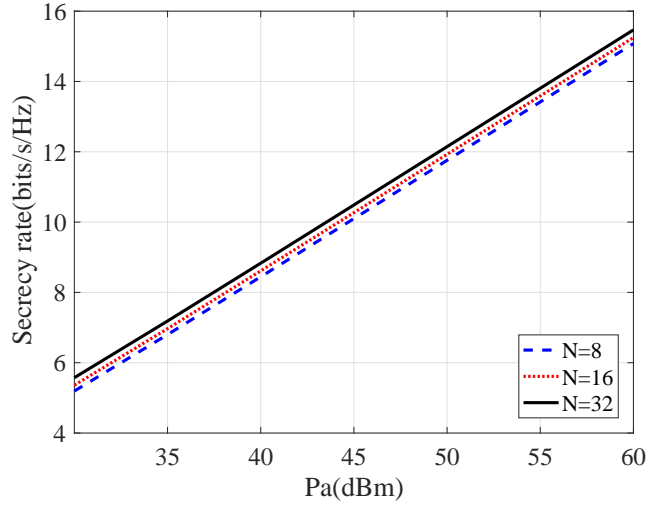}
\caption{SR versus $P_a$ for different number of antennas.}
\label{SR_n}
\end{figure}
Fig.~\ref{SR_n} illustrates the curves of SR versus $P_a$ for different antenna numbers when the UAV user operates in HD mode with the proposed strategy. It is obvious that when the transmit power is fixed, increasing the number of antennas can increase SR performance. Therefore, we can add the number of antennas to reduce the transmit power and deploy more antennas to make up for the secrecy rate.

\begin{figure}[!t]
\centering
\includegraphics[width=0.6\textwidth]{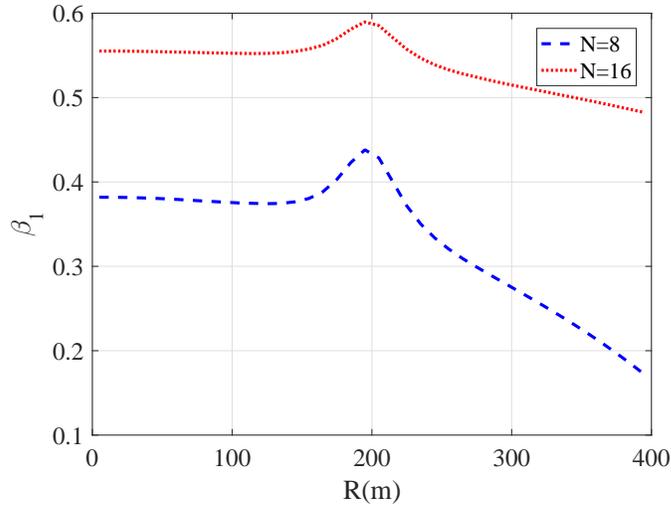}
\caption{$\beta_1$ versus distance between Alice and Bob for the proposed strategy.}
\label{beta1_r}
\end{figure}

\begin{figure}[!h]
\centering
\includegraphics[width=0.6\textwidth]{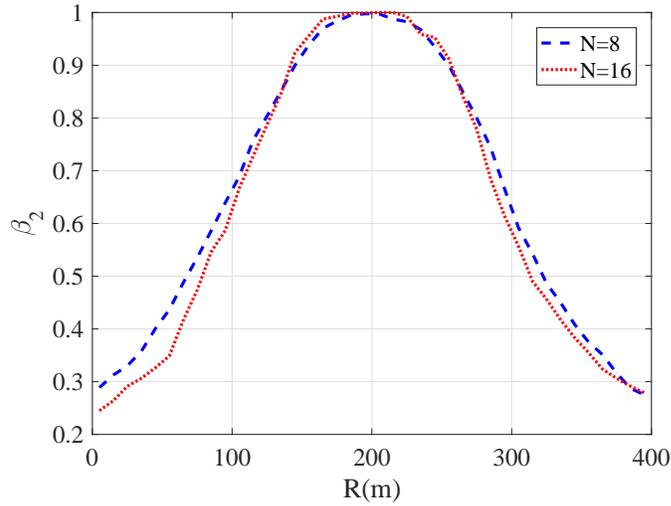}
\caption{$\beta_2$ versus distance between Alice and Bob for the proposed strategy.}
\label{beta2_r}
\end{figure}

Fig.~\ref{beta1_r} shows the curves of $\beta_1$ versus the flight distance between Alice and Bob ($R$) for the proposed strategy. It can be clearly seen that with $R$ getting close to the distance between Alice and Eve ($L_2$), $\beta_1$ begins to increase. Particularly, $\beta_1$ achieves the maximum value at $R=L_2$. On the contrary, when Bob flies farther than the distance between Alice and Eve ($R>L_2$), $\beta_1$ tends to decrease. It reveals that Alice allocates more power to transmit confidential messages when Bob flies towards Eve, and more power is used to generate AN signals when Bob flies farther away from Eve to Alice. In addition, we can obviously observe that at a fixed position, $\beta_1$ increases as the transmit antennas, which means that as the number of transmit antennas increases, the power Alice allocated to send confidential messages also grows. This accounts for the reason that deploying more antennas will result in an improvement on SR performance as shown in Fig.~\ref{SR_n}.

Compared with Fig.~\ref{beta1_r}, Fig.~\ref{beta2_r} illustrates the curves of $\beta_2$ versus the flight distance between Bob and Alice for the proposed strategy. It is seen from this figure that when Bob flies towards Eve ($R<L_2$), it increases power for transmitted AN signals to defend against the eavesdropper. And when it flies away ($R>L_2$), the AN power will slow down. All these curves disclose that the eavesdropper becomes sensitive when $|R-L_2|$ approaches zero, in which case, Alice and Bob need more power to generate AN signals for degrading the quality of the Alice-to-Eve link.

\section{Conclusion}
In this paper, we investigated a secure wireless system with a FD UAV user. To improve the SR performace, a low-complexity alternating iterative algorithm was proposed to realize a loop between the two PA factors. Firstly, in order to simplify the complicated joint optimization problem, MRT, NSP and Max-ANLNR criterion were adopted to construct the beamforming vector $\mathbf{v}_{b}$, the AN projection matrices $\mathbf{P}_{AN}$ and $\mathbf{q}_{AN}$. Then, it turned to address the bivariate PA optimization problem with the remaining two PA factors. Actually, the simplified objective function was continuous and differentiable with respect to either the PA factor $\beta_1$ or $\beta_2$. By discussing the set of critical points, we attained the optimal value of one PA factor when the other was fixed. Finally, the alternating iterative algorithm between $\beta_1$ and $\beta_2$ was proposed to further enhance the SR. Simulations shown that the PA strategy in FD mode can improve the SR performance compared with the fixed PA factors in FD mode. And it also outperformed the case where the UAV user operated in HD mode. Moreover, the SR performance of the proposed strategy grown with the increasing of the $P_a$ and the number of transmit antennas. The results disclosed that deploying more antennas and increasing the transmit power are beneficial to enhance the security of UAV-aided DM wireless system.

\Acknowledgements{This work was supported in part by the National Natural Science Foundation of China (Nos. 61771244, 61501238, 61702258, 61472190, 61801453, and 61271230), in part by the Open Research Fund of National Key Laboratory of Electromagnetic Environment, China Research Institute of Radiowave Propagation (No. 201500013), in part by the Jiangsu Provincial Science Foundation under Project BK20150786, in part by the Specially Appointed Professor Program in Jiangsu Province, 2015, in part by the Fundamental Research Funds for the Central Universities under Grant 30916011205, and in part by the open research fund of National Mobile Communications Research Laboratory, Southeast University, China (Nos. 2017D04 and 2013D02).}

\end{document}